\documentclass[%
 reprint,
superscriptaddress,
 amsmath,amssymb,
 aps,
pra,
]{revtex4-2}

\usepackage{graphicx}
\usepackage{dcolumn}
\usepackage{bm}

\begin{document}


\title{A scaling law in optomechanically induced nonlinear oscillation}

\author{Han Xiao Zhang}
 \affiliation{School of Physics and Electronic Engineering, Hainan Normal University, Haikou 571158, China\\}

 \author{Vitalie Eremeev}
 \affiliation{Instituto de Ciencias B\'{a}sicas, Facultad de Ingenier\'{i}a y Ciencias, Universidad Diego Portales, Av. Ejercito 441, Santiago, Chile}
    
\author{Jinhui Wu}
 
\affiliation{Center for Quantum Sciences and School of Physics, Northeast Normal
University, Changchun 130117, China
}%

\author{Miguel Orszag}

\author{Bing He}%
\email{bing.he@umayor.cl}
\affiliation{Multidisciplinary Center for Physics, Universidad Mayor, Camino La Pir\'{a}mide 
5750, Huechuraba, Chile 
}%

\begin{abstract}
Stable limit cycle as a stabilized mechanical oscillation is the primary result of the dynamical evolution of an optomechanical system under sufficiently powerful pump. Because this dynamical process is highly nonlinear, it was not clear whether there exists a quantitative law to relate an evolved mechanical oscillation (the limit cycle of the dynamical process) to the given parameters of the fabricated system. Here, by means of the numerical simulations based on nonlinear dynamics, we demonstrate the existence of such quantitative relations that are generally valid to the nonlinear optomechanical processes. These quantitative relations can be summarized to a scaling law that is seemingly similar to those in phase transitions of many-body systems but has very different properties. Such a quantitative law enables one to find the more feasible system parameters for realizing the same or a similar dynamical evolution result, so it will be useful to the relevant experimental researches. 
\end{abstract}

\maketitle

\section{Introduction}
Optomechanical systems (OMS) have been considered as 
a feasible platform to realize macroscopic quantumness \cite{aspelmeyer2014cavity,anetsberger2009near} and precise measurements \cite{verlot2010backaction,zhang2012precision}. Due to a unique coupling induced by the radiation pressure of its cavity field on the mechanical resonator, an OMS demonstrates the highly interested phenomena such as mechanical ground-state cooling \cite{teufel2011sideband, chan2011laser, peterson2016laser},
optomechanically induced transparency \cite{agarwal2010electromagnetically,weis2010optomechanically,safavi2011electromagnetically}, and mechanical squeezing \cite{wollman2015quantum,pirkkalainen2015squeezing}.
All these dynamical processes can be described by means of a linearization around their finally evolved states \cite{aspelmeyer2014cavity} or in an alternative approach of quantum dynamics \cite{he2017radiation, lin2014fully, lin2017mass, wang2019breaking, he2023dynamical}.

Another important category in the current research is nonlinear optomechanics, which must concern the full dynamics of an OMS. The most well-known nonlinear dynamical scenario of OMS is the Hopf bifurcation of mechanical motion to limit cycles
\cite{rokhsari2005radiation,marquardt2006dynamical,zaitsev2011forced,zhang2014self,krause2015nonlinear}, as well as the transitions into chaos \cite{carmon2007chaotic,lu2015p,monifi2016optimistically,yang2019chaotic,zhu2022cavity}, when the driving field becomes sufficiently strong. Like most other nonlinear dynamical systems, no general law in a quantitative form was known with such nonlinear dynamical processes to see how a change of the system parameters affects the system's dynamical behavior. For example, if an OMS is manufactured in a different way to have the changed system parameters, it will be usually impossible to predict whether or not the same dynamical behavior would still be preserved. On the other hand, one will be able to realize the same or similar dynamical behavior with a setup fabricated in another way or pumped by a driving field of different power, if there exists a certain law to connect the performances of two different setups. How to implement a nonlinear dynamical scenario by adjusting the system parameters flexibly to the ones that can be most conveniently realized is highly meaningful to the applications of the concerned systems.

\begin{figure}[t]
	\centering
	\includegraphics[width=\linewidth]{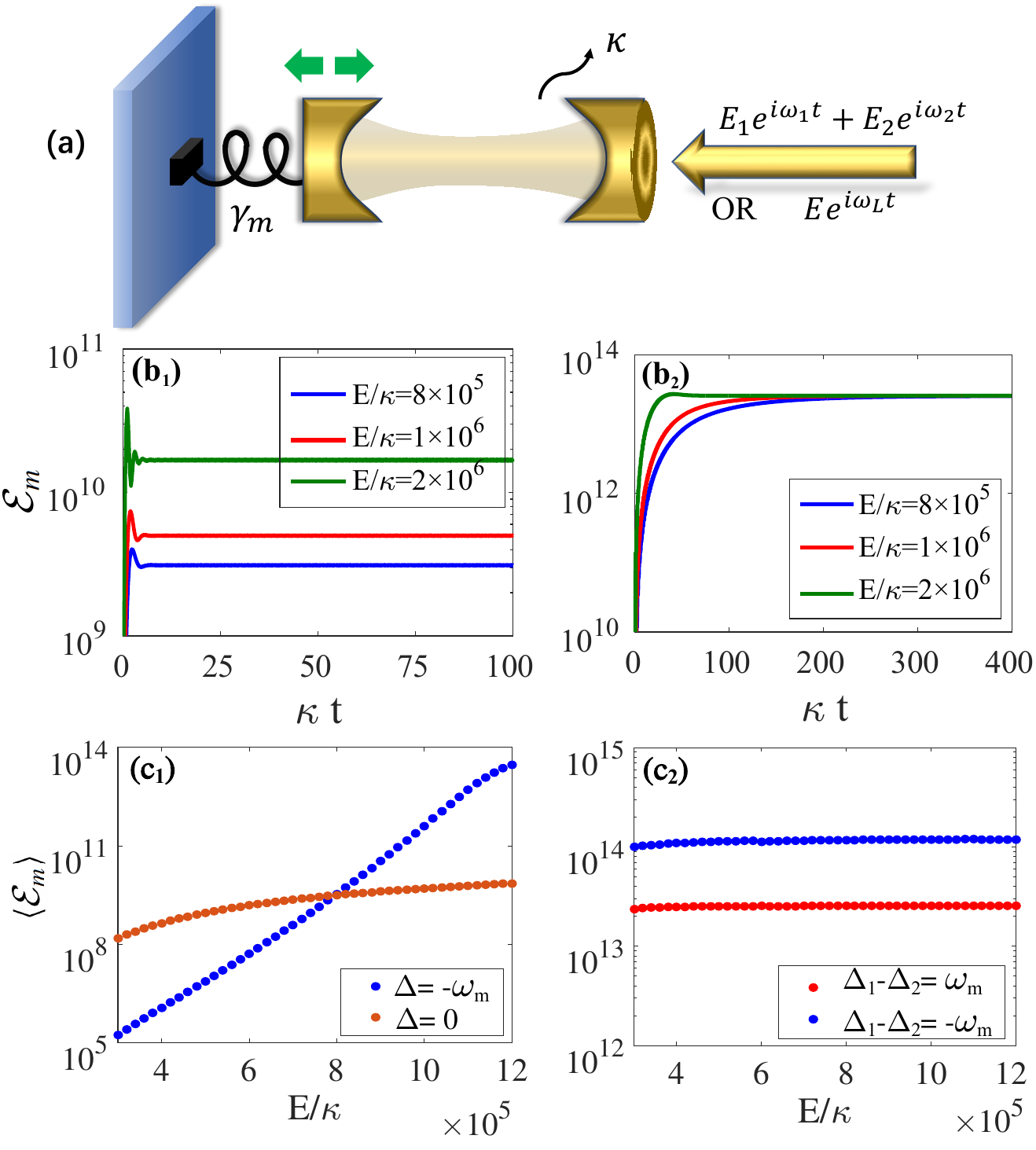}
	\caption{($a$) Generic OMS driven by a driving field of one tone or two different tones. (b$_{1}$) Sample evolution courses of the mechanical energy when the system is driven by the single-tone fields of $\Delta=-\omega_m$. (b$_{2}$) Sample evolution processes due to a driving field with one red detuned ($\Delta_{1}=\omega_m$) tone and another resonant ($\Delta_{2}=0$) tone, and they take longer time to stabilize. As in (b$_{1}$), the evolution courses are obtained at the room temperature $T=300$ K corresponding to a thermal occupation $n_{th}=1.25\times 10^5$. A stochastic function with the amplitude $\sqrt{2\gamma_m n_{th}/\kappa}$ is used to simulate the thermal noise, and the thermal occupation is based on $\omega_m=50\kappa$ with $\kappa=10^6$ Hz. Under the condition $\Delta_{1}-\Delta_{2}=\omega_m$, all evolution courses due to the different drive amplitudes $E$ converge to the same vertical position in (b$_{2}$), while it is impossible to have this behavior if any of the drive tones is removed. (c$_{1}$) and (c$_{2}$) Corresponding distributions of the evolved average $\left\langle\mathcal{E}_{m}\right\rangle$ due to the exemplary single-tone drives and two-tone drives, respectively. In (c$_{2}$), the blue one is for $\Delta_{1}=-\omega_m$ and the red one is 
	for $\Delta_{1}=\omega_m$, given the second tone as $\Delta_2=0$. Here we scale the system parameters with respect to the damping rate $\kappa$ of the associated optical cavity:  $\omega_{m}=50\kappa, g=10^{-5}\kappa$, and $\gamma_{m}=10^{-4}\kappa$.}
	\label{fig1}
\end{figure} 

In this work, we demonstrate the existence of quantitative relations in nonlinear optomechanical processes. These relations can be combined to a scaling law which specifies how a limit cycle of OMS keeps invariant under the simultaneous change of the relevant system parameters. Beyond the physical processes with an explicit self-similarity (a typical example is the study of the aerodynamics of an object by its proportional model in wind tunnel), which can be unveiled by the Buckingham $\Pi$ theorem in dimensional analysis \cite{buckingham1914physically, barenblatt1996scaling}, similar scaling laws mostly exist in the critical regimes close to the phase transitions of many-body systems \cite{widom1965equation,kadanoff1966scaling} or near the bifurcations of other dynamical systems (in terms of varied mathematical forms) \cite{leonel2021scaling}. Concerning a nonlinear system without any explicit symmetry in its dynamical equations, our illustrated scaling law relates its dynamical evolution process to the parameters determined by the fabrication of the system, and it works without being restricted to the regimes of bifurcation. 

We will illustrate the scaling relations through the dynamical scenarios summarized in Fig. 1(a): a system of optomechanics stabilizes 
in the oscillations of its coupled cavity field and mechanical oscillator, under either a pump laser of one frequency or a driving field with two different frequencies. Under one particular condition for the two-tone pump field, the induced mechanical oscillation will have its oscillation totally locked without changing its amplitude, even if the pump power is varied over a considerable range \cite{he2020mechanical, wu2022amplitude}. The above-mentioned scaling law is valid to both scenarios of single-tone and two-tone drives, which are being currently explored in the field of optomechanics. 

The rest of the paper is organized as follows. In Sec. II we first present the main features of the concerned systems and, then, state the scaling law for their dynamical processes by a specific example. 
In Sec. III we take the examples of OMS under a single-tone drive to illustrate the details about how to find the explicit relations between the stabilized mechanical oscillations and the relevant system parameters. These relations are further studied with the systems under two-tone drives in Sec. IV. Before the concluding section, Sec. V is focused on the applications of the discovered scaling relations to experimental research, together with the illustration of another scaling relation for the realization of locked mechanical oscillations under two-tone driving fields.

\section{A law of dynamics for driven cavity optomechanical processes}

We here describe a general optomechanical dynamics process in the reference frame rotating at the resonant frequency $\omega_c=cN\pi/L$ ($c$ is the speed of light, $N$ an integer, and $L$ the cavity size) of the cavity, so that the detuning $\Delta=\omega_c-\omega_{l}$ of the driving laser frequency $\omega_l$ from the resonant cavity frequency $\omega_c$ becomes a relevant parameter to write the system Hamiltonian 
of a single-tone drive as (also see the derivation with a two-tone field in Appendix) 
\begin{align}
H&=(1/2)\hbar\omega_m(X_m^2+P_m^2)-\hbar gX_m|a|^2\nonumber\\
&+i\hbar E(a^\dagger e^{i\Delta t}-a e^{-i\Delta t}).
\label{system}
\end{align}
The dimensionless mechanical displacement and momentum used in the above are related to the real ones as $x_m=\sqrt{\frac{\hbar}{m\omega_m}}X_m$ and $p_m=\sqrt{m\hbar\omega_m}P_m$ ($\omega_m$ is the mechanical frequency and $m$ the effective mass of the mechanical resonator). Given the intracavity field with a photon number $|a|^2$, the coefficient $g$ is the optomechanical coupling constant at the single-photon level. 

After including the associated dissipation-fluctuation effects with the damping rate $\kappa$ ($\gamma_{m}$) of the cavity (mechanical oscillator), one has the dynamical equations
\begin{align}
\dot{a} & =-\kappa a+igX_ma+Ee^{i\Delta t},\nonumber\\
\dot{X}_m &  =\omega_{m}P_{m},\nonumber\\
\dot{P}_m &  =-\omega_{m}X_{m}-\gamma_{m}P_{m}+g|a|^2+\sqrt{2\gamma_m}\xi_m(t),
\label{eq}
\end{align}
with the thermal noise term $\xi_m(t)$ 
satisfying the correlation $\langle \xi_m(t)\xi_m(t')\rangle= (2n_{th}+1)\delta(t-t')$ ($n_{th}$ is the thermal occupation at a certain temperature), because this noise can be well approximated as a white noise at room temperature \cite{lin2017mass,lin2020entangling}. Meanwhile, the cavity field noise is negligible here due to an effective zero temperature of the cavity field reservoir (as compared with the large $n_{th}$ of the mechanical noise). When the cavity field damping at the rate $\kappa$ is dominated by the pump-cavity coupling (the intrinsic loss is negligible to a high-quality cavity), the drive amplitude $E$ is related to the pump powers $P$ 
as $E=\sqrt{\frac{2\kappa P}{\hbar\omega_l}}$. For the dynamical processes like the one described by Eq. (\ref{eq}), we will mathematically scale the evolution time to the dimensionless one $\kappa t$, and the used parameters will be correspondingly scaled to $g/\kappa$, $\omega_{m}/\kappa$, and $\gamma_m/\kappa$, so that the evolution results are more general than those of a specific optical cavity.

\begin{figure}[h]
	\centering\includegraphics[width=\linewidth]{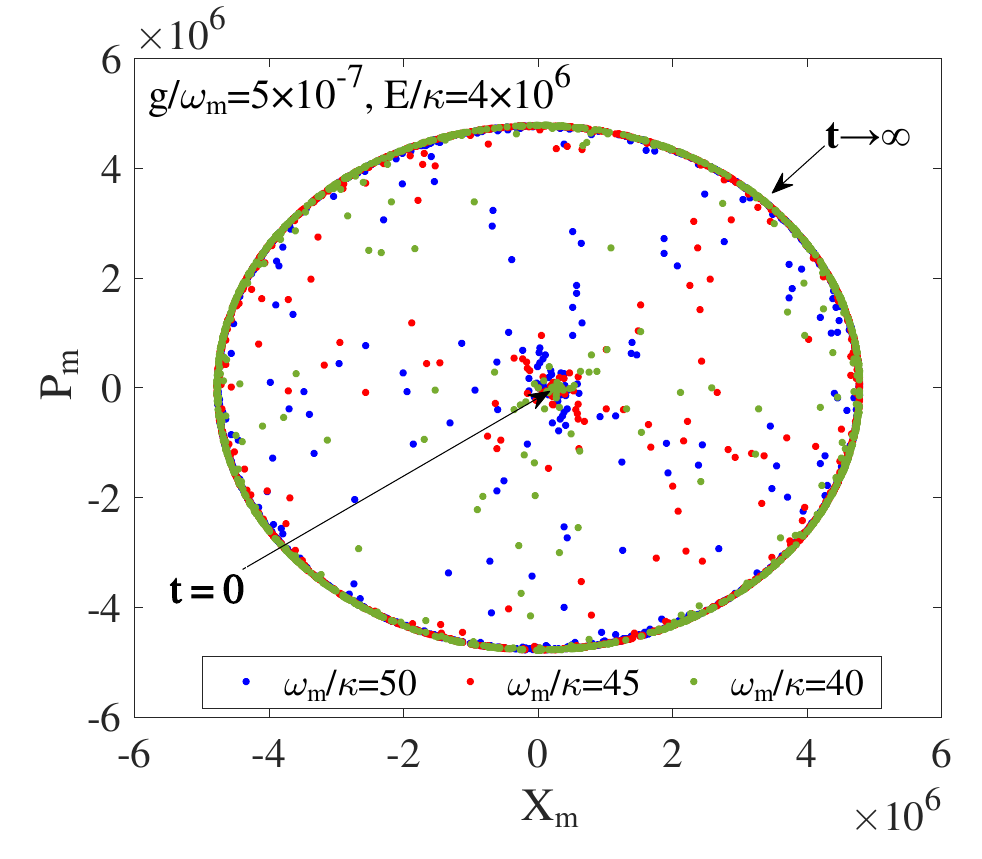}
	\caption{Exemplary evolution processes for different systems, which are depicted in the phase space ($X_m,P_m$). The three systems with different mechanical oscillation frequencies $\omega_m$ have the varied optomechanical coupling constants $g$, but share the same ratio $g/\omega_{m}$. Then, under the same pump power or the same pump amplitude $E$ of a single-tone drive, these systems will evolve from the position $(X_m,P_m)=(0,0)$ at $t=0$ to the same limit cycle as $t\rightarrow\infty$. The pump detuning is $\Delta=-\omega_m$ and the mechanical damping rate is chosen as $\gamma_{m}=10^{-5}\kappa$. }
	\label{fig2}
\end{figure}

We will focus on the mechanical part of an OMS. For the scenario of single-tone drive in Eq. (\ref{eq}), the pump power $P$ should be over the threshold of a Hopf bifurcation to have the finally evolved mechanical motion branched from a static equilibrium to a stable oscillation as limit cycle. Some sample processes are illustrated in terms of the real-time mechanical energy
\begin{align}
	\mathcal{E}_m(t)&=\frac{1}{2}X_m^2(t)+\frac{1}{2}P_m^2(t)
	\label{em}
\end{align}
as in Figs. 1(b$_1$) and 1(b$_2$), respectively, for the single-tone and two-tone drives. The main point of the current research is the existence of a scaling behavior demonstrated in Fig. \ref{fig2}: the same limit cycle for the different dynamical systems will be approached, if the first two terms of the Hamiltonian in Eq. (\ref{system}) change together by the same scale so that the ratio $g/\omega_{m}$ is invariant, while the third term of the Hamiltonian keeps unchanged.

In practice, it is more convenient to illustrate the associated scaling relations in terms of the averaged dimensionless mechanical energy
\begin{align}
	\langle \mathcal{E}_m\rangle&=\frac{1}{2}\langle X_m^2\rangle+\frac{1}{2}\langle P_m^2\rangle
\end{align}
over one mechanical oscillation period, which is equivalent to an phonon number of the mechanical mode. This quantity, which can be approximated as $\langle \mathcal{E}_m\rangle\approx (1/2)A^2$ for a stabilized mechanical oscillation $X_m(t)=A\sin(\Omega_mt+\phi)+d$ ($d\ll A$ and $\Omega_m\neq \omega_m$ due to optical spring effect), 
is in one-to-one correspondence with the stable limit cycle of mechanical motion, since it is proportional to the area of the evolved limit cycle. Then, the behavior of limit cycles in Fig. 2 can be formulated as the following scaling law:
\begin{align}
	\left\langle \mathcal{E}_{m}\right\rangle=(E/\kappa)^{2}\cdot F(gE/\omega_m),
	\label{scale2}
\end{align}
where we define $F$ as a dimensionless scaling function with a variable $gE/\omega_m$ of the unit Hz. The evolved average mechanical energy $\langle \mathcal{E}_m\rangle$ due to the single-tone and two-tone drives are very different as seen from Fig. 1(c$_1$) and Fig. 1(c$_2$), but they follow the same scaling law, Eq. (\ref{scale2}). We will explain how to find this nontrivial scaling law, which cannot be obtained directly from Eq. (\ref{eq}).

\section{Unraveling the scaling behaviors with single-tone drives}

The exemplary dynamical processes in Fig. 1 are according to Eq. (\ref{eq}) but with a set of fixed system parameters, $g$, $\omega_m$, $E$ and others. It is natural to ask this question---how such evolved mechanical energy will be changed if the system parameters are varied? The answer is not straightforward because Eq. (\ref{eq}) is highly nonlinear. Since the mechanical resonator in an OMS is under 
two forces, the radiation pressure and spring restoring force, one can estimate that the evolved mechanical energy is determined by these three parameters, $g$, $E$, and $\omega_{m}$. 

By observations, the stabilized $\left\langle \mathcal{E}_{m}\right\rangle$ of a nonlinear dynamical process is determined by the interplays of the above-mentioned parameters. One such interplay is that the effect of a lower nonlinear magnitude $g$ can be compensated by a stronger pump power to increase the cavity photon number, so that what is relevant to the nonlinear dynamical process is their product $gE$.
Another point is that the evolved mechanical energy should vanish if the pump laser is turned off, i.e., the evolved energy $\left\langle \mathcal{E}_{m}\right\rangle$ must be zero given $E=0$. This asymptotic behavior requires a prefactor $(E/\kappa)^{\beta}$ for the energy $\left\langle \mathcal{E}_{m}\right\rangle$, where $\beta$ could be a real number. Combining these considerations together, we present the first ansatz: 
\begin{align}
	\left\langle \mathcal{E}_{m}\right\rangle=(E/\kappa)^{\beta}\cdot f(gE/\kappa^2,\omega_m/\kappa)
	\label{scale}
\end{align}
in terms of the function $f(gE/\kappa^2,\omega_m/\kappa)$ with two dimensionless variables.

This ansatz can be numerically verified. In Fig. 3(a), we especially adopt the evolution processes of anomalous stabilization \cite{zhang2024highly}, i.e., there exists a transition between two different metastable states (a sudden jump) during their evolution courses, to establish one property of Eq. (\ref{scale}). The reason is that it is clearer to demonstrate the similarity of the two trajectories in Fig. 3(a); with the vertical axis in the logarithmic scale, they can be displaced to each other along the vertical axis, because their jumps during the evolution processes occur at the same time. When the system of $g=10^{-4}\kappa$ is replaced by another one fabricated to have a lower coupling constant $g=10^{-5}\kappa$, 
a completely similar evolution trajectory of $\mathcal{E}_{m}(t)$ can be realized by increasing the drive amplitude $E$, while the product $gE$ keeps to be the same, though a high drive amplitude $E$ will lead to a higher $\left\langle \mathcal{E}_{m}\right\rangle$ in the end.  
A striking feature for such finally stabilized values of the energy $\left\langle \mathcal{E}_{m}\right\rangle$, which are due to the simultaneously changed $g$ and $E$ but under the condition of a fixed product $gE$, is that their distribution is highly regular. Under the conditions $gE=constant$ and $\omega_m=constant$, we plot such values of $\ln\left\langle \mathcal{E}_{m}\right\rangle$ against the quantity $\ln (E/\kappa)$ in Fig. 3(b), finding their linear relations with an exact slope $\beta=2$. This particular value of $\beta=2$ for the prefactor of the ansatz Eq. (\ref{scale}) is general to the nonlinear dynamical processes of OMS. It should be noted that the linear relation in Fig. 3(b) is valid only after the systems enter their stable oscillations, i.e., above their corresponding Hopf bifurcation points. 

\begin{figure}[t]
	\centering\includegraphics[width=\linewidth]{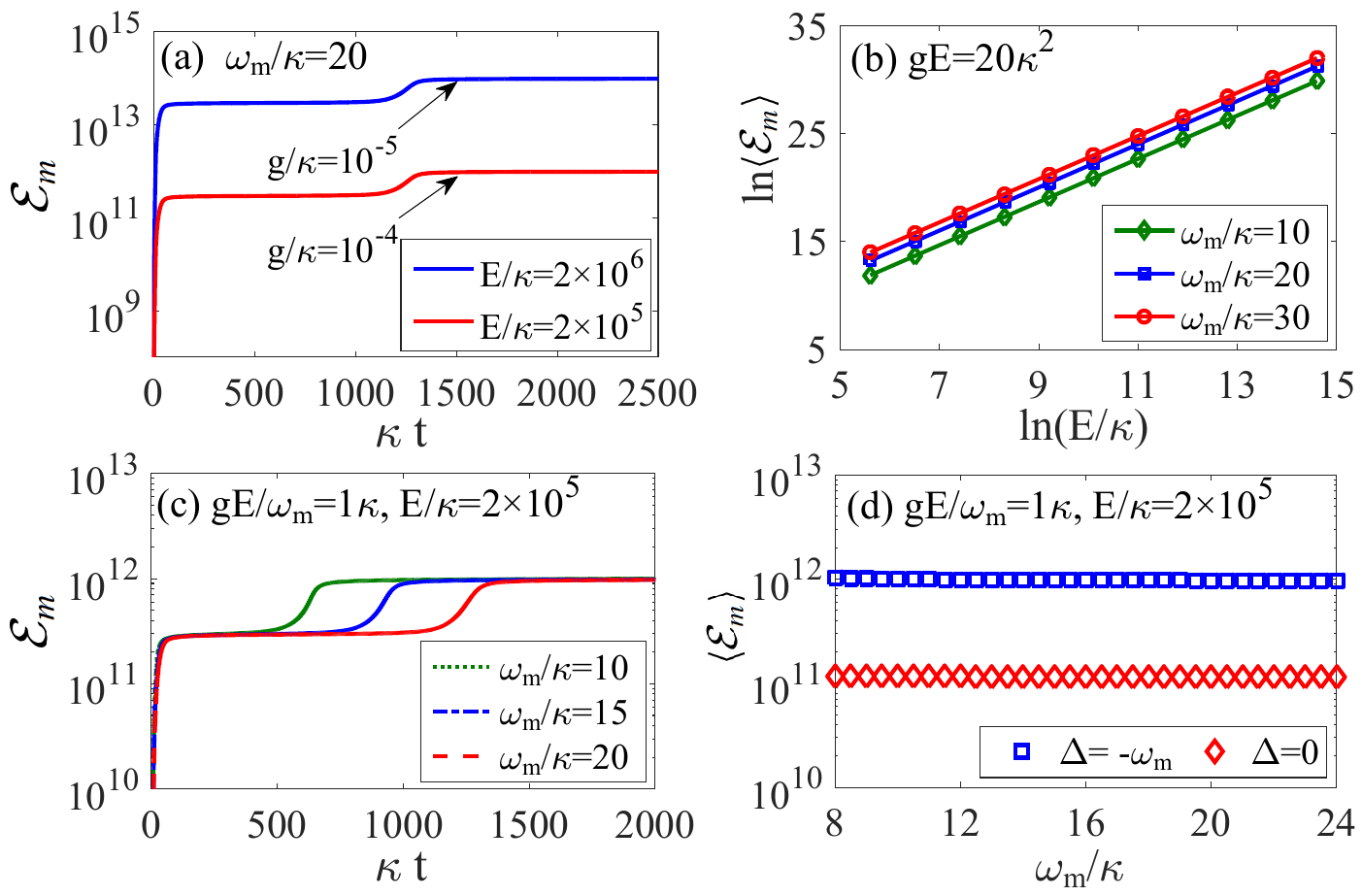}
	\caption{(a) Completely similar dynamical evolution processes under the condition that $gE$ is constant. (b) Corresponding linear relations between the finally evolved $\ln\left\langle \mathcal{E}_{m}\right\rangle$ and $\ln (E/\kappa)$, for the systems with their different mechanical frequencies $\omega_{m}$. Their slopes are  exactly found to be $2$. (c) Sample dynamical evolution processes for the systems with the same value of $gE/\omega_{m}$ and under the same pump amplitude $E$. (d)  Corresponding distributions of the finally evolved $\left\langle \mathcal{E}_{m}\right\rangle$ under the conditions $gE/\omega_{m}=constant$ and $E=constant$. In (a), (b) and (c), the pump fields are blue-detuned with $\Delta=-\omega_{m}$, while two indicated values of detuning are used in (d). The mechanical damping rate is chosen as $\gamma_{m}=10^{-5}\kappa$. }
	\label{fig3}
\end{figure}

Another property of the evolved mechanical energy $\left\langle \mathcal{E}_{m}\right\rangle$ comes from the observation that the radiation pressure is balanced by the spring restoring force proportional to the term $-\omega^2_mX_m$. Then, one would speculate that the quantity $gE/\omega_m$ partially determines the concerned dynamical behaviors, since a stronger restoring force due to a higher $\omega_m$ neutralizes the effect of more radiation pressure growing with the product $gE$. Based on this assumption, it is expected that the evolved average energy $\left\langle \mathcal{E}_{m}\right\rangle$ will be the same given a fixed quantity $gE/\omega_m$ for different systems, together with the same drive amplitude $E$ considering the prefactor in Eq. (\ref{scale}). 

The above estimation can be confirmed by the dynamical evolution processes shown in Figs. 3(c), where, if one varies the optomechanical coupling $g$ together with the mechanical frequency $\omega_m$, the evolving trajectories under the same drive amplitude $E$ will nonetheless tend to the same final value of $\mathcal{E}_{m}$. Here, the trajectories display their jumps at different moments of time, due to their different intrinsic mechanical frequencies $\omega_{m}$. Particularly, once an OMS stabilizes to a mechanical oscillation, all others sharing the same value of $gE/\omega_m$ and under the same pump power can reach the same final oscillation too. This fact is further proved by the examples in Fig. 3(d), where one sees that the mechanical energy values realized under a fixed ratio $gE/\omega_m$, together with a constant drive amplitude $E$, always locate on the same horizontal level along the axis of the parameter $\omega_m$. These results allow one to reformulate Eq. (\ref{scale}) to the scaling law in Eq. (\ref{scale2}), which well predicts the dynamical behaviors of the concerned nonlinear systems.

The scaling law in Eq. (\ref{scale2}) is physically nontrivial as compared with an explicit self-similarity of nonlinear system, which can be seen by means of the Buckingham 
$\Pi$-theorem \cite{buckingham1914physically,barenblatt1996scaling}. Note that all system parameters we adopt are with the same unit Hz ($s^{-1}$); see Eqs. (\ref{system}) and (\ref{eq}). 
The relation about the system parameter combinations in Eq. (\ref{scale2}) manifests a deeper physical meaning: all realized mechanical oscillations come into being from a competition between the radiation pressure and the spring restoring force. On the other hand, the energy $\left\langle \mathcal{E}_{m}\right\rangle$ in Eq. (\ref{scale2}) is different from a free energy of two variables, $f(x,y)$, in a critical regime of phase transition, which is a homogeneous one such that $f(\lambda x,\lambda y)=\lambda^\gamma f(x,y)$ ($\gamma$ is a real number) and can be reduced to the form $f(x,y)=x^\gamma \phi(y/x)$ \cite{widom1965equation,kadanoff1966scaling}. While it can exist far away from a bifurcation point, the energy $\left\langle \mathcal{E}_{m}\right\rangle$ is obviously not a homogeneous function of $gE$ and $\omega_m$; otherwise the pre-factor $E^2/\kappa^2$ before the scaling function in Eq. (\ref{scale2}) will become $(gE/\kappa^2)^2$.

\section{Scaling relations of the locked oscillations under two-tone pumps}

Next, we will present the corresponding scaling relations for the systems under two-tone drives, with the first equation of Eq. (\ref{eq}) being replaced by
\begin{align}
	\dot{a} & =-\kappa a+igX_ma+E(e^{i\Delta_1 t}+e^{i\Delta_2 t}).
	\end{align}
So far, two-tone driven optomechanical process has been experimentally studied with one tone of blue detuned ($\Delta_{1}=-\omega_{m}$) and another tone of red detuned ($\Delta_2=\omega_m$) \cite{shomroni2019two}. Under another condition that the difference $|\Delta_1-\Delta_2|$ of the two drive tones is close to the mechanical frequency $\omega_m$, an OMS pumped by such two-tone field will enter a special dynamical pattern, in which the stabilized mechanical oscillation,
\begin{align}
X_m(t)=A\sin(\omega_mt+\phi)+d,
\label{oscillation}
\end{align}
has the almost unchanged amplitude $A$, frequency $\omega_{m}$ (this frequency is locked to be the one determined by the system fabrication), as well as the fixed phase $\phi$, irrespective of the applied drive power that can be adjusted over a considerable range \cite{he2020mechanical, wu2022amplitude}. Here we assume two equal drive amplitudes at $E$, though the unequal pump tones can also lead to the similar phenomena \cite{wu2022amplitude}. The relevant feature to our concerned scaling relations is the locking of the amplitude $A$ so that the evolved mechanical energy becomes almost invariant with the increased drive amplitude $E$ as in Fig. 1(c$_2$). After the drive power or the corresponding drive amplitude $E$ is increased much further, the mechanical oscillation can be locked to other orbits with higher amplitudes $A$, which are also determined by the fabrication of a system \cite{he2020mechanical}. The locked mechanical oscillations have been proposed to the applications in precise measurement \cite{li2022ultra, yan2023optomechanical, yan2023force}. It is also possible for the single-tone drives to achieve an amplitude locking \cite{zhang2024highly}, but it requires much higher pump power. For example, to a system of $g=10^{-5}\kappa$ and $\omega_{m}=10\kappa$, the locking to the lowest mechanical oscillation amplitude can be realized with the two-tone drive amplitudes $E$ in the order of $10^4\kappa$, in contrast to the required $E\sim 10^6\kappa$ under the corresponding single-tone drives with $\Delta=-\omega_{m}$. Moreover, the frequency locking is impossible for the single-tone scenario, i.e., there exists an unavoidable mechanical frequency shift from the value $\omega_{m}$ determined by the system fabrication \cite{zhang2024highly}.

\begin{figure}[t]
	\centering\includegraphics[width=\linewidth]{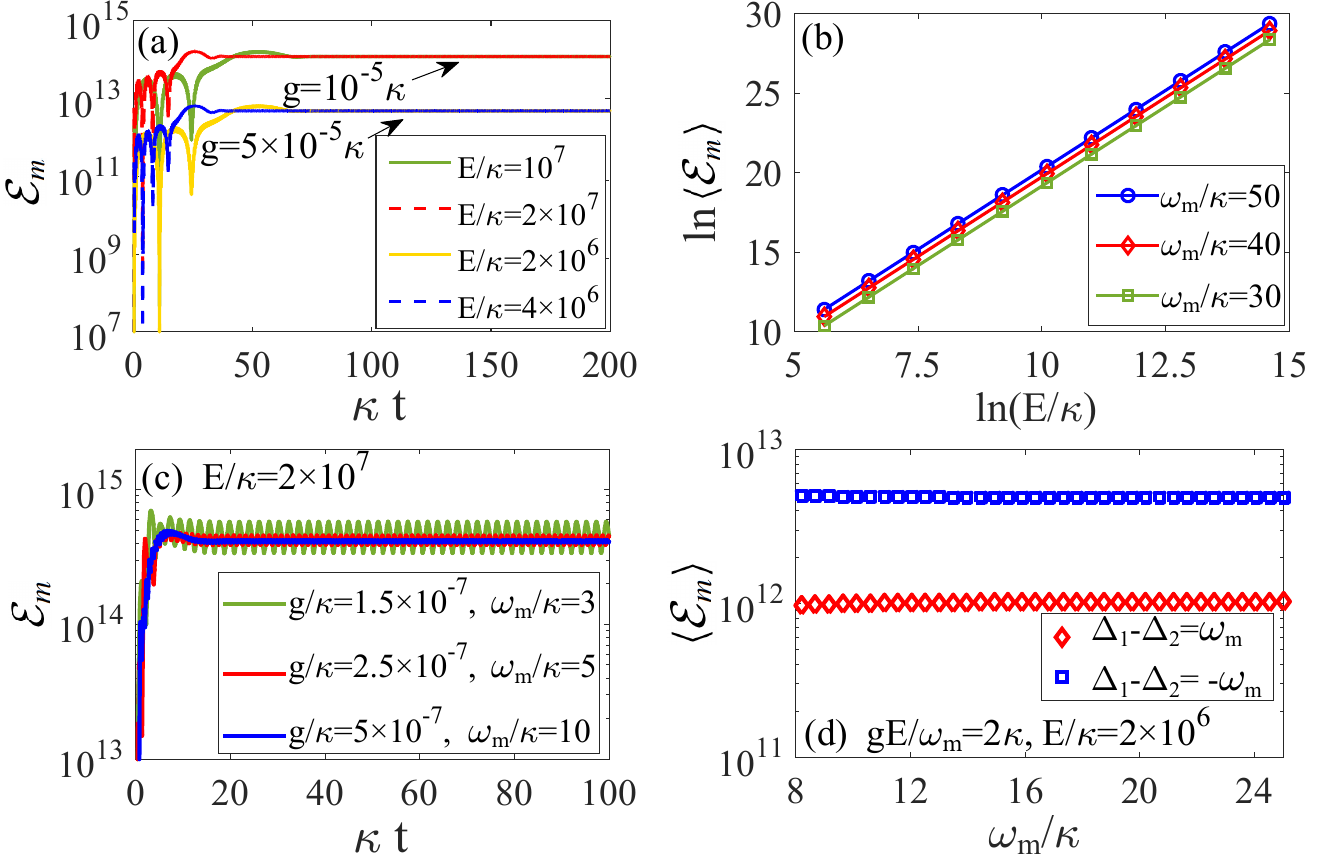}
	\caption{(a) Completely similar dynamical evolution processes under the condition that $gE$ is constant. (b) Corresponding linear relations between the finally evolved $\ln\left\langle \mathcal{E}_{m}\right\rangle$ and $\ln (E/\kappa)$ for the systems with their different mechanical frequencies $\omega_{m}$. Their slopes are  exactly found to be $2$. (c) Sample dynamical evolution processes for the systems with the same value of $gE/\omega_{m}$ and under the same pump amplitude $E$. (d) Corresponding distributions of the finally evolved $\left\langle \mathcal{E}_{m}\right\rangle$ with $\omega_{m}$, under the conditions $gE/\omega_{m}=constant$ and $E=constant$. In (a) and (b), one pump tone is blue detuned with $\Delta_{1}=-\omega_{m}$ and the other is resonant as $\Delta_2=0$. In (c) the combination of the drive tones 
		is  with $\Delta_{1}=\omega_{m}$ and  $\Delta_{2}=0$. In (d) the blue line is due to one blue-detuned tone with $\Delta_1=-\omega_{m}$ and the red line is due to one red-detuned tone $\Delta_1=\omega_{m}$, while the other tone is a resonant one in both cases. The mechanical damping rate is chosen as $\gamma_{m}=10^{-4}\kappa$. }
	\label{fig4}
\end{figure}

In Fig. 4(a), we respectively choose two groups of drive amplitudes $E$ such that any of the systems with their different $g$ evolves to the same oscillation under both amplitudes $E$ of the same group, like what is described in Fig. 1(b$_2$). With one group of the drive amplitudes differed by two times, e.g., $E=2\times 10^6\kappa$ and $4\times 10^6\kappa$, the trajectories of real-time mechanical energy $\mathcal{E}_{m}(t)$ for the system of $g=5\times 10^{-5}\kappa$ approach the same mechanical energy in the end. For another system of the lower optomechanical coupling $g=10^{-5}\kappa$, two exactly similar trajectories can be obtained after the corresponding drive amplitudes $E$ are increased to the values of the other group that preserve the products $gE$; compare the two pairs of trajectories in Fig. 4(a), including their transient periods before stabilization. As exemplified in Fig. 4(b), all stabilized $\left\langle \mathcal{E}_{m}\right\rangle$ under the condition $gE=constant$ satisfy a linear relation, for any fixed mechanical frequency $\omega_m$. The straight lines in Fig. 4(b) have the exact slope of $\beta=2$, the exponent of the prefactor on the right-hand side of Eq. (\ref{scale2}). 

The locked mechanical oscillations realized by the different drive amplitudes $E$ applied to the same system have almost the same mechanical energy $\left\langle \mathcal{E}_{m}\right\rangle$. Experimentally one will observe the same sideband spectrum of the cavity fields \cite{he2020mechanical}, which is characteristic of a locked oscillation (the respective amplitudes of the spectrum components are proportionally changed by the differed drive amplitudes $E$ so that the extra energy due to a higher $E$ predominantly stays in the cavity field). If one changes the optomechanical coupling $g$, together with its intrinsic mechanical frequency $\omega_{m}$, so that their ratio $g/\omega_{m}$ is fixed, the finally realized average mechanical energy without varying the applied pump power will not be changed. This fact can be seen from the evolution processes of the real-time mechanical energy $\mathcal{E}_{m}(t)$ under such condition; see the examples in Fig. 4(c). The only difference of these evolved trajectories under the same ratio $g/\omega_m$ and the same drive amplitude $E$ is in the oscillation amplitudes of the real-time energy
\begin{align}
	\mathcal{E}_{m}(t)=\frac{1}{2}A^2+\frac{1}{2}d^2+Ad\sin(\omega_mt+\phi),
	\label{energy}
\end{align}
corresponding to Eq. (\ref{oscillation}) with $d\ll A$. Under the two-tone drives, the pure displacements $d$ are relatively large due to a higher static force from  the field intensity $|a|^2$, and then the amplitude $Ad$ of the real-time energy $\mathcal{E}_{m}(t)$ (the amplitude $A$ is fixed for the locked oscillation) is magnified to be more obvious. Against a smaller restoring force due to a lower $\omega_m$, there is a larger displacement $d$ under the radiation pressure, thus giving rise to a larger amplitude $Ad$ of $\mathcal{E}_{m}(t)$ in Fig. 4(c). A more straightforward view of the invariant mechanical energy $\left\langle \mathcal{E}_{m}\right\rangle$ under the fixed $gE/\omega_m$ and $E$ is presented in Fig. 4(d), where we consider two different combinations of the drive tones. This scaling relation in Fig. 4(d) will be violated (the horizontal lines in Fig. 4(d) will be broken), only when the systems are locked to the higher orbits of oscillation that are easily influenced by random perturbations \cite{he2020mechanical}.  

\begin{figure}[t]
	\centering\includegraphics[width=\linewidth]{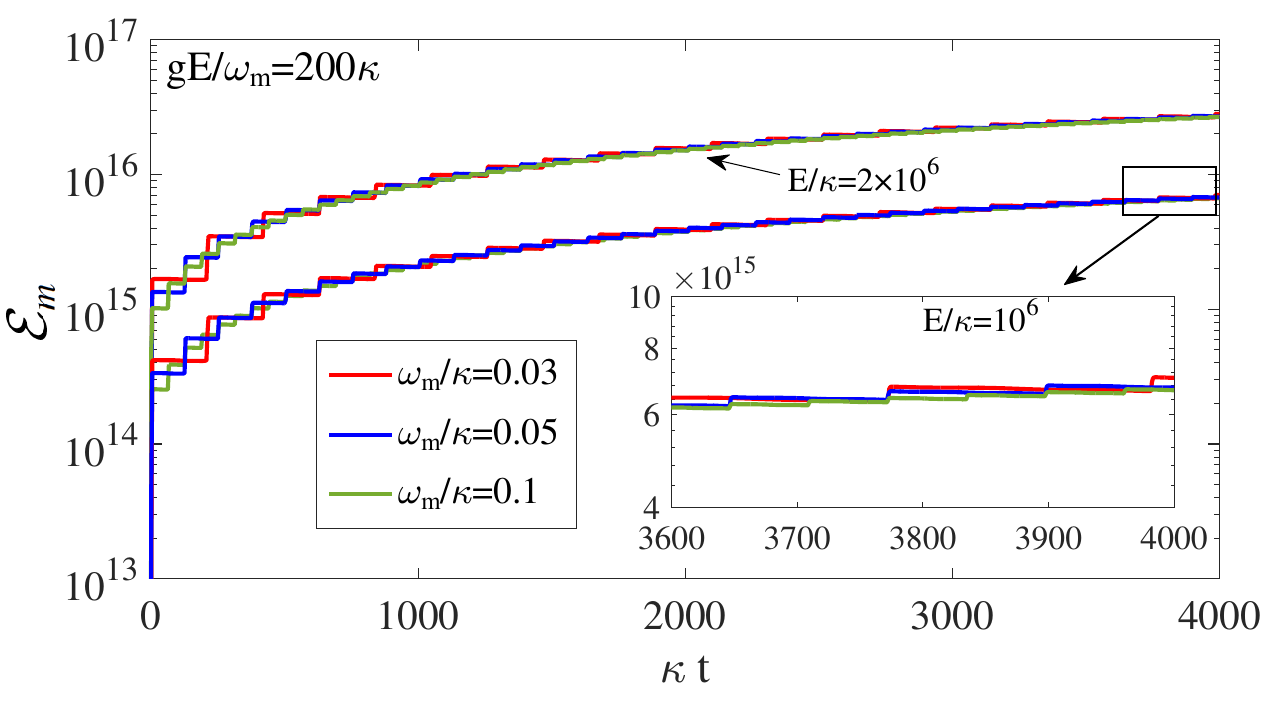}
	\caption{Scaling behavior of the dynamical evolution under a two-tone drive satisfying $|\Delta_1-\Delta_2|=\omega_m$ but in the regime $\omega_m<\kappa$. Here, the step-like jump of the mechanical energy occurs after each mechanical oscillation period due to a push by the pulsed cavity field. Under the indicated conditions, the mechanical energy trajectories for the systems with different $\omega_m$ but 
	the same $\gamma_{m}=10^{-4}\kappa$ approximately evolve together in the end, having the same time-dependent scaling function $F(gE/\omega_m,t)$ for Eq. (\ref{scale2}). }
	\label{fig5}
\end{figure}

A different dynamical behavior will come into being when the ratio $\omega_m/\kappa$ is lowered to be less than one, so that the system enters the so-called unresolved sideband regime. 
The associated cavity field of this scenario can turn into a pulsed one \cite{poot2012backaction,gao2015self,miri2018optomechanical,xu2021chip}.
If one applies two drive tones satisfying $|\Delta_1-\Delta_2|=\omega_m$ to the systems of $\omega_m/\kappa<1$, it will enter an instability due to a correlated pulsed field with the mechanical oscillation \cite{lin2023nonlinear}; in each mechanical oscillation period, the field pulse acting opposite to the mechanical resonator's motion will be suppressed so that the mechanical amplitude will first become higher and higher, and then stabilizes due to the intrinsic damping rate $\gamma_{m}$ after very long period of time \cite{gu2024optical}. The step-like mechanical energy $\mathcal{E}_{m}(t)$ induced by pulses also satisfies the scaling law in Eq. (\ref{scale2}), as demonstrated in Fig. \ref{fig5} where the mechanical energy trajectories under the conditions, $g/\omega_m=constant$ and $E=constant$, finally proceed almost together for the systems of various $\omega_{m}$.  

It is possible to develop the scenario in Fig. \ref{fig5} into a generator of optical frequency comb \cite{gu2024optical}. The key point for having a broad band of this type of frequency combs is the quantity $gA/\omega_m$ ($A$ is the dimensionless mechanical oscillation amplitude), which should be as large as possible in the realized optomechanical oscillation. Even for a rather high dimensionless mechanical energy $\mathcal{E}_{m}\sim 10^{16}$ in Fig. \ref{fig5}, the corresponding real mechanical displacements $x_m=\sqrt{\frac{\hbar}{m\omega_m}}X_m$ are within the order of $10$ nm, given the microresonators used for a previous experimental demonstration of optomechanical frequency combs \cite{hu2021generation}, which have a zero-point fluctuation amplitude $\sqrt{\frac{\hbar}{m\omega_m}}\sim 0.1$ fm. Such mechanical displacements are still much less than the size $L\approx 30$ $\mu$m of the microresonators so that the next-order correction to the Hamiltonian in Eq. (1) can be neglected. In the previous experiment with single-tone pumps \cite{hu2021generation}, the ratio $gA/\omega_m\sim gA/\kappa$ (the total cavity damping rate $\kappa$ is tuned close to $\omega_{m}$) has nearly reached $1000$, though the associated mechanical oscillation amplitudes $A$ are smaller than those in Fig. 5. The scaling law in Eq. (\ref{scale2}) can help to find the optimum system parameters for this type of frequency-comb generators, as in the designs of other dynamical processes discussed in the next section.   

\section{Experimental Applications}

The linear relation in Fig. 3(b) and Fig. 4(b) states that the similar optomechanical oscillations can be realized over a large span of the drive amplitude, e.g., $e^{5}\kappa<E<e^{15}\kappa$. Then, there is flexibility to choose the proper $g$ and $E$ for the implementation of optomechanical oscillations. However, any currently available OMS has a small optomechanical coupling constant $g$, and the upper bound for the possibly applied drive amplitude $E$ is also limited due to the possible damage by a strong pump. According to Eq. (\ref{scale2}), therefore, it is beneficial to select a mechanical frequency $\omega_m$ as small as possible, but keeping the same ratio $gE/\omega_m$ at the same time (so that the product $gE$ can be lowered accordingly), for the excitation of an optomechanical oscillation. In other words, the systems should not be deep in the resolved sideband regime ($\omega_m\gg \kappa$) required for optomechanical cooling \cite{aspelmeyer2014cavity, teufel2011sideband, chan2011laser, peterson2016laser}. 

Beyond the above qualitative consideration, the more important role of the scaling law in Eq. (5) is to provide the quantitative guides for obtaining similar dynamical processes. Different dynamical scenarios, such as those under single-tone drives and under two-tone drives satisfying the condition $|\Delta_{1}-\Delta_{2}|=\omega_{m}$, have different scaling functions $F(gE/\omega_{m})$, but these functions or the corresponding functions $f(gE/\kappa^2,\omega_m/\kappa)$ in Eq. (\ref{scale}) are of the same form to each individual category of dynamical processes. Then, one can well design the implementation of an optomechanical oscillation with a certain mechanical amplitude $A$ or its corresponding $\left\langle \mathcal{E}_{m}\right\rangle$ by searching the optimum values in the space $(gE/\kappa^2,\omega_m/\kappa)$, which are the closest to the available system fabrication. A particular example is the oscillation locking under two pump tones satisfying the conditions $|\Delta_1-\Delta_2|=\omega_m$ and $\omega_m/\kappa>1$. Any driven OMS with the ratio $gE/\omega_m=1.0\kappa$ surely reaches such locked oscillation. Given a realizable ratio $\omega_m/\kappa=5$ and a cavity damping rate $\kappa=2\pi\times 10^6$ Hz, for example, the required power for the drive tones near the wavelenghth $\lambda=1537$ nm can be reduced from the level of watt to the order of milliwatt if the coupling constant $g$ can be increased from $1$ Hz to $10$ Hz. The required system parameters can be more relaxed with an even smaller ratio $gE/\omega_m$ that can realize a locked optomechanical oscillation. 
 
\begin{figure}[h]
	\centering\includegraphics[width=\linewidth]{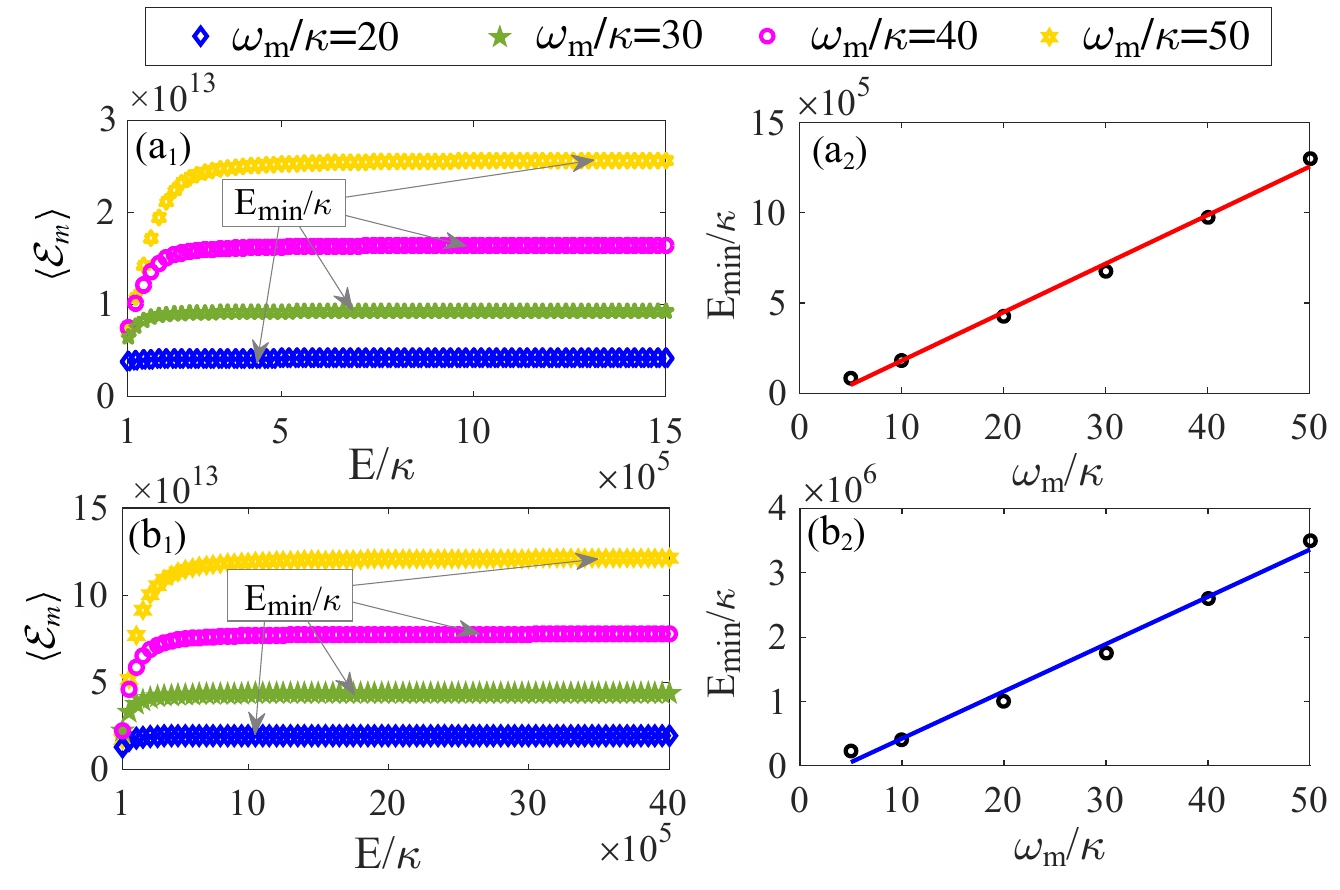}
	\caption{(a$_{1}$) Samples of the minimum drive amplitudes for the systems with a number of varied $\omega_m$ to enter a locked oscillation. Here, one of the drive is red-detuned and the other is resonant with the cavity frequency $\omega_c$. (a$_{2}$) Linear relation between $E_{min}$ and $\omega_m$; the fitted line is $E_{min}/\kappa=2.694\times 10^{4}(\omega_m/\kappa)-8.929\times10^{4}$. (b$_{1}$) and (b$_{2}$) Corresponding results after one drive tone is replaced to be blue-detuned. The fitted line is $E_{min}/\kappa=7.335\times 10^{4}(\omega_m/\kappa)-3.097\times10^{5}$ in (b$_{2}$). In these figures we fix $g=10^{-5}\kappa$ and $\gamma_{m}=10^{-4}\kappa$ (these parameters are experimentally achievable). 
	The appearance of the tendencies in Figs. 6(a$_1$) and 6(b$_1$) differs from the one in Fig. 1(c$_2$) because a normal scale (instead of the logarithmic scale) is used here.}
	\label{fig6}
	\vspace{-0cm}
\end{figure}
 
To an experimental realization of the locked oscillation under the two-tone drives satisfying the condition $|\Delta_{1}-\Delta_{2}|=\omega_{m}$, the most interested issue is how large the minimum drive amplitude $E_{\min}$ (or the corresponding minimum pump laser power) is required to enter the regime of oscillation locking. This question leads to the finding of another scaling relation. The process of entering the locked mechanical energy with the increased $E$ is gradual without a specific bifurcation point \cite{he2020mechanical} (this process is not so obvious in Fig. 1(c$_2$) due to the logarithmic 
scale used there), so there is a freedom to say where is the minimum $E_{\min}$. We here define the point $E_{\min}$ with the derivative $\kappa\frac{d\langle \mathcal{E}_{m}\rangle}{dE}= 10^5$ (this quantity is chosen to be dimensionless); the system is regarded as being on the locked oscillation completely when any $E>E_{\min}$ gives the derivative $\kappa\frac{d\langle \mathcal{E}_{m}\rangle}{dE}$ less than this value. This choice of the $E_{\min}$, which can be adjusted to a different $E$ in the nearby, is illustrated in Fig. \ref{fig6}($a_{1}$) and Fig. \ref{fig6}($b_{1}$), respectively, for both situations involving one red-detuned or one blue-detuend drive tone. The distribution of its associated mechanical energy is due to the scaling function $F(gE/\omega_{m})\sim(gE/\omega_m)^{-2}$ on the locked orbits, so that the corresponding mechanical energy takes the form $\left\langle \mathcal{E}_{m}\right\rangle\sim (\omega_m/g)^2$, the flat parts of the curves in Figs. \ref{fig6}($a_{1}$) and \ref{fig6}($b_{1}$). At the minimum $E_{\min}$, the locked oscillation for the system with the parameters, $\omega_m=10\kappa$ $g=10^{-5}\kappa$ and $\gamma_{m}=10^{-4}\kappa$, has $\left\langle \mathcal{E}_{m}\right\rangle=1.038 \times 10^{12}$ under one red-detuned and another resonant tone, and it will become $\left\langle \mathcal{E}_{m}\right\rangle=4.995 \times 10^{12}$ if the red detuned tone is replaced by a blue-detuned one. It is seen that, for a larger $\omega_m$, the required $E_{\min}$ to reach the locked mechanical oscillation should be higher. Plotting the obtained minimum drive amplitudes $E_{\min}$ against $\omega_m$ gives the linear relation [see Figs. \ref{fig6}($a_{2}$) and \ref{fig6}($b_{2}$)]:
\begin{align}
E_{\min}/\kappa=c\left(\omega_m/\kappa\right)+d,
\label{scale3}
\end{align}
where $c>0$. Similar linear relations exist if the definition of the minimum $E_{\min}$ is slightly modified to other values of $\kappa\frac{d\langle \mathcal{E}_{m}\rangle}{dE}$. By the extension of the linear relation in Fig. \ref{fig6}($a_{2}$) to an experimentally feasible point at $\omega_m=5\kappa$, the required minimum $E_{min}$ will be lowered to $4.54\times 10^{4}\kappa$. Given an experimental OMS reported in Ref. \cite{hu2021generation} ($\omega_{c}=2\pi\times 194 $ THz, $\omega_m=2\pi\times 50.1$ MHz, $g/\sqrt{\frac{\hbar}{m\omega_m}}=2\pi\times 6.35$ GHz/nm, and an intrinsic loss rate of the cavity at $\kappa_i=2\pi\times 1.7$ MHz), it corresponds to a pump power about $9.98$ mW.

\section{Conclusion}
In the past, the influence of system parameters on the evolution of a dynamical system was mainly studied in the theory of bifurcations \cite{kuznetsov1998elements,hale2012dynamics}. Here we explore a different type of relations between the evolution results and the parameters of a dynamical system described by Eq. (\ref{eq}), which is possibly generalizable to other driven systems with their dynamical equations containing quadratic nonlinear terms. We have found the explicit scaling behaviors in the inherent dynamics of OMS. They can be summarized to the scaling law in Eq. (\ref{scale2}), which states that the same limit cycle for the mechanical motion can be approached under the simultaneous change of the optomechanical coupling $g$ and mechanical frequency $\omega_m$ to keep their ratio, while the size of the 
limit cycle is determined by the pump laser power. In other words, one will see a nontrivial behavior that the same limit cycle will be reached if the first two terms of the Hamiltonian in Eq. (1) are scaled together but its third drive term keeps invariant. According to Eq. (\ref{scale2}), the different behaviors of various types of dynamical scenario, such as the processes under single-tone drives or those of two-tone drives satisfying a proper condition for their drive tones, are mainly manifested by the different scaling functions $F(gE/\omega_m$) which, however, take the same form 
to each category of dynamical processes. Such a general law can be applied to the designs of the relevant experiment setups. Many decades ago, the scaling relations discovered for many-body systems \cite{widom1965equation,kadanoff1966scaling} significantly facilitated the understanding of phase transition phenomena. The scaling relations that have been numerically verified for a few-body nonlinear system may as well provide the better understandings 
of the concerned nonlinear dynamical processes. 

\vspace{-0cm}
\renewcommand{\theequation}{A-\arabic{equation}}
  \setcounter{equation}{0}  
  \section*{APPENDIX}
Under a driving field with two tones at the frequencies $\omega_1$ and $\omega_2$, respectively, the total Hamiltonian of a cavity optomechanical system, which is modeled by two nonlinearly coupled oscillators, is given as
\begin{align}
H_T&=(1/2)\hbar\omega_m(\hat{X}_m^2+\hat{P}_m^2)+\hbar\omega_c \hat{a}^{\dagger}\hat{a}-\hbar g\hat{X}_m\hat{a}^{\dagger}\hat{a}\nonumber\\
&+i\hbar \sum_{k=1}^2E_k(\hat{a}^\dagger e^{-i\omega_{k}t}-\hat{a} e^{i\omega_{k}t}).
\label{system0}
\end{align}
The interaction term comes from the first order correction of the cavity frequency under the radiation pressure, due to the fact that the mechanical displacements $\sqrt{\frac{\hbar}{m\omega_m}}X_m$ are always much less than the cavity sizes of the realistic setups. Then, one has the following dynamical equations of the cavity-field and mechanical parts from the above Hamiltonian:
\begin{eqnarray}
\dot{a}&=&-\kappa a-i(\omega_c-g_mX_m)a+E_1e^{-i\omega_1 t}+E_2e^{-i\omega_2 t},\nonumber\\
\dot{X}_m &  =&\omega_{m}P_{m},\nonumber\\
\dot{P}_m &  =&-\omega_{m}X_{m}-\gamma_{m}P_{m}+g|a|^2.
\label{1}
\end{eqnarray}
These equations are the mean-field approximations of the corresponding Heisenberg-Langevin equations (note that the dynamical variables are reduced to the c-numbers without hats).
By a transformation $a\rightarrow ae^{-i\omega_c t}$ of the cavity field mode, the first equation is transformed to 
\begin{eqnarray}
&&\dot{a}=-\kappa a+ig_mX_ma+E_1e^{i\Delta_1 t}+E_2e^{i\Delta_2 t}.
\label{2}
\end{eqnarray}
Now, the corresponding Hamiltonian after the transformation is the one that is similar to the form in Eq. (1). 
Another form of Hamiltonian carries an extra term proportional to $\Delta \hat{a}^{\dagger}\hat{a}$ from a different rotation by the drive frequency \cite{aspelmeyer2014cavity}. For the scenario of two-tone drives, it is more convenient to simplify the equations by a rotation with respect to the cavity resonance frequency $\omega_c$.

\vspace{-0cm}
\section*{Acknowledgements}
Supported by ANID Fondecyt Regular (Grant No. 1221250) and Hainan Provincial Natural Science Foundation of China (Grant No. 122QN302).

\raggedright  
\bibliography{scaling-law_oms}

\end{document}